\documentclass{article}[14pt]
\usepackage[english]{babel}
\usepackage{epsfig}
\usepackage{amssymb}
\usepackage{amsmath}

\begin{document}

\author{B. R\'{o}\.{z}ycki$^{1,2}$ and M. Napi\'{o}rkowski$^{1}$ \\
  $^{1}$ Instytut Fizyki Teoretycznej, Uniwersytet Warszawski \\
  00-681 Warszawa, Ho\.za 69, Poland \\
  $^{2}$Max-Planck-Institut f\"ur Kolloid- und Grenzfl\"achenforschung, \\
  14424 Potsdam, Germany }

\title{Phase Transitions in Multicomponent String Model}

\maketitle

\begin{abstract}
  We propose a one-dimensional model of a string decorated with
  adhesion molecules (stickers) to mimic multicomponent membranes in
  restricted geometries. The string is bounded by two parallel walls
  and it interacts with one of them by short range attractive forces
  while the stickers are attracted by the other wall. The exact
  solution of the model in the case of infinite wall separation
  predicts both continuous and discontinuous transitions between
  phases characterised by low and high concentration of stickers on
  the string. Our model exhibits also coexistence of these two phases,
  similarly to models of multicomponent membranes. \\

\noindent PACS numbers: 05.70.Fh, 05.70.Np, 87.16.Dg
\end{abstract}

The functional renormalization group calculations showed that the
continuous unbinding transitions of fluid membranes are characterised
by the same fixed points as those of one-dimensional strings governed
by a finite tension \cite{RL88}. However, one-dimensional strings
attracted to a flat substrate by finite-range forces exhibit only
continuous unbinding transitions while two-dimensional fluid membranes
can undergo also discontinuous unbinding transitions in this case
\cite{RL94}. Moreover, recent computer simulations indicated that
models for multicomponent membranes show complex phase diagrams which
contain both continuous and discontinuous unbinding transitions
\cite{TWRL01}. The aim of this letter is to present a model for
two-component, one-dimensional string fluctuating between two parallel
walls with which it interacts. This model is solved exactly in the
case of infinite wall separation and exhibits phase transitions
similar to multicomponent membranes.

Any membrane present in biological cells is composed of a lipid
bilayer which provides its basic structure, and which contains
different proteins \cite{biol}. The proteins anchored in the lipid
bilayer are large molecules which can usually freely diffuse within
the fluid membrane. These macromolecules protrude from the bilayer and
may interact with another membrane, or a substrate, by a short-range
forces and, thus, act as local stickers or repellers. To investigate
the influence of this interaction on membrane adhesion phenomenon
various models for membranes with stickers in contact with a planar
substrate were used recently \cite{TWRL01, TW02, 4}. In these models
a membrane with zero spontaneous curvature is usually considered and
adhesion molecules of only one sort are taken into account. The state
of the membrane is specified then by two fields: the local separation
between the membrane and the substrate and the local concentration of
the stickers. As already mentioned, both continuous and discontinuous
membrane unbinding transitions were found for such models, and
separation into the unbound phase with low average concentration of
stickers and the bound phase with higher average sticker concentration
was observed.

\begin{figure}
\begin{center}
\includegraphics[scale=0.7]{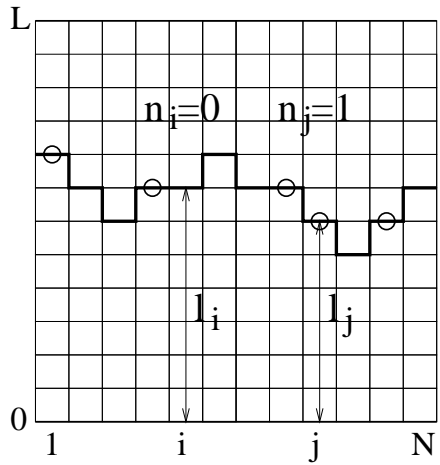}
\end{center}
\caption{One-dimensional lattice model for a string with stickers. Position of the string at site number $i$ is denoted by $l_{i}$, and $n_{j}=1$ (or 0) indicates presence (or absence) of a sticker at the $j$-th site. }
\label{lattice}
\end{figure}

In this letter we consider a model of a string containing specific
adhesion molecules (stickers). The string with attached stickers
fluctuates in a two-dimensional space between two parallel,
one-dimensional walls (slit geometry).  The walls are separated by
distance $L$. In the present lattice model (see Fig. \ref{lattice})
the string separation from the bottom wall is described by the set of
discrete variables $l_{i}$, where the lattice site number $i=1, \ldots
,N$ measures the distance along the one-dimensional walls. Each of the
$N$ variables $l_{i}$ can take $L+1$ values: $l_{i}=0, \ldots ,L$. The
stickers' degrees of freedom are specified by the set of occupation
numbers $n_{i}$ and each of these variables can be either $0$ or $1$.
The value $n_{i}=1$ indicates the presence of a sticker on the string
above the i-th lattice site while the value $n_{i}=0$ indicates that
there is no sticker at that position.

In the present model we assume that the string itself interacts with 
the upper wall via a
pinning potential while the stickers attached
to the string interact only with the bottom wall by another pinning
potential. The corresponding Hamiltonian takes the following form 
\begin{equation}
H= \sum_{i=1}^{N} \left[ J|l_{i+1}-l_{i}|
-n_{i} W_{1} \delta_{l_{i},0} - W_{2} \delta_{l_{i},L} \right] \quad.
\label{1}
\end{equation}
The first term describes the bending energy of the string, and
$J$ is the stiffness parameter. The second and the third terms
correspond to the interaction of the stickers with the bottom wall and
the interaction of the string with the upper wall, respectively; we
assume that $W_{1}>W_{2}>0$. 

The grand partition function
\begin{equation}
Z(T, \mu, N)= \sum_{\left\{ n_{i} \right\}} \sum_{\left\{ l_{i} \right\}} 
\exp \left[- \beta \left( H - \mu \sum_{i=1}^{N} n_{i} \right) \right]
\quad,
\end{equation}
where $\beta = \frac{1}{k_{B}T}$ and $\mu$ denotes the chemical
potential of the stickers, will be evaluated for periodic boundary
conditions along the walls. To this aim we first sum out the stickers
degrees of freedom $\left\{ n_{i} \right\}$, and then - using transfer
matrix method - trace out all possible string configurations $\left\{
l_{i} \right\}$. We additionally assume that the string does not
fluctuate too violently, and so its positions at the neighbouring
sites differ at most by one, i.e., $|l_{i+1}-l_{i}|=0,1$. In this way
we arrive at the modification of the restricted solid-on-solid model
\cite{RSOS}.  This leads to the following exact result
\begin{equation}
Z(T, \mu, N) = \left( 1+ e^{\beta \mu} \right)^{N} 
\sum_{i=0}^{L} (\lambda_{i})^{N}
\quad,
\label{2}
\end{equation}
where the quantities $\lambda_{0} \ge \lambda_{1} \ge \ldots \ge
\lambda_{L}$ are the eigenvalues of the corresponding transfer matrix.
It is given explicitly by the following expression
\begin{equation}
\mathbb{T}_{m,n}=
\left( \delta_{m,n}+ j \delta_{|m-n|,1} \right)
\sqrt{ 
w_{1}^{\delta_{m,0} + \delta_{n,0}}
w_{2}^{\delta_{m,L} + \delta_{n,L}}
} ,
\label{3}
\end{equation}
where
\begin{equation}
j=e^{-\beta J},
\quad 
w_{1}=\frac{1+e^{\beta(W_{1}+\mu)}}{1+e^{\beta \mu}}, 
\quad
w_{2}=e^{\beta W_{2}}
\quad,
\label{4}
\end{equation}
and $\delta_{i,j}$ is Kronecker delta. A transfer matrix of similar
structure was discussed in \cite{BRMN} in the context of the wetting
phenomena in the slit geometry. 

In the thermodynamic limit $N \to \infty$ the free-energy density
\begin{equation}
f = - k_{B}T \log  [ \lambda_{0} (1+e^{\beta \mu}) ]
\quad,
\label{7a}
\end{equation}
and the parallel correlation length of the string 
\begin{equation}
\xi_{||} = \left[ \log \left( 
\frac{\lambda_{0}}{\lambda_{1}}
\right) \right]^{-1}
\quad,
\label{7b}
\end{equation}
are determined by the two largest eigenvalues of the transfer matrix. 
The average sticker concentration $n$ can be obtained as the derivative of 
the free-energy density $f$ with respect to the chemical potential
$\mu$,
\begin{equation}
n = - \frac{\partial f}{\partial \mu}
\quad.
\label{8}
\end{equation}

The unbinding transition, by definition, occurs when the average
separation of a membrane from a substrate becomes infinite. The
discussion of this phenomenon within the present model is only
possible if the distance between the walls is made infinite, $L =
\infty$. In this limiting case the spectrum of the transfer matrix
$\mathbb{T}$ consists of infinite number of eigenvalues homogeneously
distributed in the segment $]1-2j, 1+2j[$, and - in addition - of at
most two eigenvalues which can be larger than $1+2j$. We arrive at
this result using the same methods as discussed in detail in
\cite{BRMN}. The two largest eigenvalues of the transfer matrix are
expressed in terms of two auxiliary parameters $\sigma_{1}$ and
$\sigma_{2}$
\begin{equation}
\lambda_{0}= \max \left[ \sigma_{1}, \sigma_{2} \right]
\quad, \quad \lambda_{1}= \min \left[ \sigma_{1}, \sigma_{2} \right]
\label{9}
\end{equation}
where
\begin{equation}
\sigma_{k}= \left\{
\begin{array}{lll}
\frac{w_{k}}{2} \left[ 1+ \sqrt{1+ \frac{4j^{2}}{w_{k}-1}} \right] 
& \rm{for} & w_{k} \ge \frac{1+2j}{1+j} \\
1+2j & \rm{for} & w_{k} < \frac{1+2j}{1+j}
\end{array}
\right.
\quad,
\label{10}
\end{equation}
for $k=1,2$. The parameters $\sigma_{k}$ have continuous first derivatives 
with respect to $w_{k}$, and for $w_{k} \ge \frac{1+2j}{1+j}$ one has 
$\sigma_{k} \ge 1+2j$, and $\frac{\partial \sigma_{k}}{\partial w_{k}} \ge 0$.

The unbinding phase transition is indicated by divergence of the
parallel correlation length (\ref{7b}), when the two largest
eigenvalues of the transfer matrix become equal, $\lambda_{0}=
\lambda_{1}$. Upon changing the model parameters the two largest
eigenvalues $\lambda_{0}$ and $\lambda_{1}$ may both become equal
either (i) to the value $1+2j$, which leads to the continuous
unbinding transitions, or (ii) to the value located inside the segment
$]1+2j,+\infty[$, and then the string unbinds discontinuously from one
of the walls and becomes pinned to the other one; these scenarios were
discussed in \cite{BRMN}. Thus the discontinuous unbinding transitions
caused by energetic competition between the infinitely separated walls
occur when $w_{1}=w_{2}>\frac{1+2j}{1+j}$, while the continuous
de-pinning transitions driven by the string thermal fluctuations take
place when $w_{1}=\frac{1+2j}{1+j}>w_{2}$ or
$w_{2}=\frac{1+2j}{1+j}>w_{1}$, see Eq. (\ref{10}). A straightforward
analysis of the above inequalities -- based on definitions (\ref{4})
-- leads to the conclusion that unbinding of the string from the
bottom wall is either the discontinuous transition, which occurs when
$\mu = \mu_{I}(T)$ and $\mu_{I}(T)>\mu_{II}(T)$, or the continuous
transition, which takes place when $\mu=\mu_{II}(T)$ and
$\mu_{I}(T)<\mu_{II}(T)$, where
\begin{equation}
\mu_{I}(T)= - \frac{1}{\beta} \log 
\frac{e^{\beta W_{1}} - e^{\beta W_{2}}}{e^{\beta W_{2}} - 1}
\quad,
\label{11}
\end{equation}
and
\begin{equation}
\mu_{II}(T)= - \frac{1}{\beta} \log \left[
\left( e^{\beta J}+1 \right) \left( e^{\beta W_{1}}-1 \right) -1
\right]
\quad.
\label{12}
\end{equation}
The tricritical point is determined by equation
$\mu_{I}(T)=\mu_{II}(T)$ and the corresponding temperature will be
denoted as $T_{2}$.
\begin{figure}
\begin{center}
\includegraphics[scale=0.7]{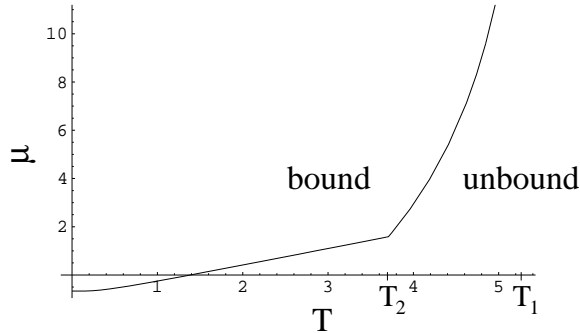} 
\end{center}
\caption{Phase diagram in coordinates $T$ - $\mu$ plotted for fixed values of model parameters: $W_{1}=2J$, $W_{2}=\frac{4}{3}J$. The chemical potential $\mu$ is given in energy units of the stiffness parameter $J$ and the temperature $T$ is given in units of $J/k_{B}$. Unbinding of the string form the bottom wall is discontinuous for low temperatures ($T<T_{2}$) and continuous for high temperatures ($T_{2}<T<T_{1}$). } 
\label{T-mu}
\end{figure}

The corresponding phase diagram plotted in the variables $T$ and $\mu$
(for fixed values of model parameters $J$, $W_{1}$, and $W_{2}$) which
contains the relevant parts of curves $\mu=\mu_{I}(T)$ and
$\mu=\mu_{II}(T)$ is displayed in Fig. \ref{T-mu}. The area above the
curve corresponds to the states for which the string is bound to the
bottom wall. The area under the curve corresponds to the states for
which the average distance between the string and the bottom wall is
infinite. For temperatures $T<T_{2}$ the curve represents the locus of
discontinuous phase transition while for $T>T_{2}$ it represents the
continuous transitions. The continuous transition line has the
vertical asymptote $T=T_{1}$, which is determined by condition
$\mu_{II}= \infty$.

To construct the phase diagram in coordinates $T$ - $n$ we have to
make use of Eq. (\ref{8}) which together with Eqs (\ref{7a}) and
(\ref{3}) leads to the following relation
\begin{equation}
n = \frac{1}{1+e^{- \beta \mu}} + 
\frac{e^{-\beta \mu}(e^{\beta W_{1}}-1)}{(1+e^{-\beta \mu})^2}
\frac{1}{\lambda_{0}} 
\frac{\partial \lambda_{0}}{\partial w_{1}} 
\quad.
\label{13}
\end{equation}
The derivative $\frac{\partial \lambda_{0}}{\partial w_{1}}$ is not
equal to zero only if $\lambda_{0}= \sigma_{1} >1+2j$, i.e., when the
string is bound to the bottom wall. In all other cases $\frac{\partial
  \lambda_{0}}{\partial w_{1}} = 0$ and then the average sticker
concentration is equal $(1+e^{-\beta \mu})^{-1}$. It means that the
configurations when the string is bound on the bottom wall correspond
to the phase with high concentration of stickers, while configurations
where the average distance of the string from the bottom wall is
infinite correspond to the phase with low sticker concentration.

The continuous unbinding transitions line $\mu_{II}(T)$ is determined
by equations $w_{1}=\frac{1+2j}{1+j}$ and $\lambda_{0}= \sigma_{1}$.
When this conditions are fulfilled then it follows from Eq.
(\ref{10}) that $\frac{\partial \lambda_{0}}{\partial w_{1}}=0$. Thus,
upon crossing the continuous transitions line $\mu_{II}(T)$ the
average sticker concentration changes continuously.

On the other hand, the discontinuous unbinding transitions take place
when $w_{1}> \frac{1+2j}{1+j}$. In this case $\frac{ \partial
  \lambda_{0}}{\partial w_{1}} > 0$ for $\lambda_{0}= \sigma_{1}$, and
$\frac{ \partial \lambda_{0}}{\partial w_{1}} = 0$ for $\lambda_{0} =
\sigma_{2}$, which means that upon crossing the discontinuous
unbinding transitions line, $\mu_{I}(T)$, the average stickers
concentration changes discontinuously.

The above predictions can be made more explicit by evaluating the
derivative $\frac{\partial \lambda_{0}}{\partial w_{1}}$ along both
transitions lines. In this way one finds the stickers concentration
(\ref{13}) on both sides of curves $\mu_{I}(T)$ and $\mu_{II}(T)$. A
straightforward calculation leads to the conclusion, that along the
continuous transitions line, on both sides of the curve $\mu_{II}(T)$,
the average stickers concentration is given by the following
expression
\begin{equation}
n_{II}(T)= \frac{1}{(e^{\beta J}+1)(e^{\beta W_{1}}-1)}
\quad.
\end{equation}
On the other hand, the average stickers concentration along the discontinuous
transitions line, $\mu_{I}(T)$, on the side of the phase poor in
stickers, is equal
\begin{equation}
n_{I}^{-}(T)= \frac{e^{\beta W_{2}}-1}{e^{\beta W_{1}}-1}
\quad,
\end{equation}
and on the side of the phase rich in stickers it is equal
\begin{equation}
n_{I}^{+}(T)= n_{I}^{-}(T) + \Delta n(T)
\quad,
\end{equation}
where the jump of stickers concentration, $\Delta n$, is given
by formula
\begin{equation}
\Delta n (T)= 
\frac{e^{\beta W_{1}}-e^{\beta W_{2}}}{2(e^{\beta W_{1}}-1)}
\left[1-2e^{- \beta W_{2}} + \left( 
1+ \frac{4e^{-2 \beta J}}{e^{\beta W_{2}}-1} \right)^{- \frac{1}{2}} \right]
\ge 0 \quad.
\end{equation}

One can check that $\Delta n (T_{2})=0$, and show that the derivative
$\left( \frac{\partial}{\partial T} \Delta n \right)_{T=T_{2}}$ is
always negative.  It means that while approaching the tricritical
point the jump of the sticker concentration vanishes linearly with
temperature, $\Delta n \sim T_{2}-T$.

\begin{figure}
\begin{center}
\includegraphics[scale=0.7]{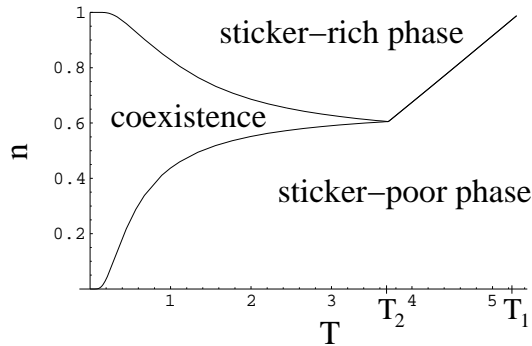} 
\end{center}
\caption{The phase diagram in coordinates $T$ - $n$ for fixed values of model parameters: $W_{1}=2J$ and $W_{2}=\frac{4}{3}J$. The temperature $T$ is given in units of $J/k_{B}$. The lines $n_{II}(T)$, $n_{I}^{-}(T)$, and $n_{I}^{+}(T)$ separate the sticker-rich phase, the sticker-poor phase, and the coexistence region of these two phases.}
\label{T-n}
\end{figure} 

An example of phase diagram, which contains the relevant parts of
curves $n_{I}^{-}(T)$, $n_{I}^{+}(T)$, and $n_{II}(T)$, is shown in
figure \ref{T-n}. The curve $n_{II}(T)$ separates the regimes which
correspond to phases rich and poor in stickers. The area limited by
curves $n_{I}^{-}(T)$, $n_{I}^{+}(T)$ and the $n$ axis is the
coexistence region of these two phases.

In summary, by analogy with some models for multicomponent membranes,
we consider a one-dimensional model of a string with stickers attached
to it. The string is confined between two parallel walls of different
properties: the bottom wall interacts only with the stickers, and the
upper wall only with the string itself. In the case when the distance
between the walls is infinite, the string unbinding transitions take
place for certain values of temperature and the sticker chemical
potential. At low temperatures ($T<T_{2}$) the transitions are
discontinuous while for higher temperatures ($T_{2}>T>T_{1}$) they are
continuous.  We show that the configurations for which the string is
bound to the bottom wall correspond to the phase with high
concentration of stickers while configurations where the average
separation between this wall and the string is infinite correspond to
the phase with low sticker concentration. We find both discontinuous
and continuous transitions between phases rich and poor in stickers.
The phase diagram on Fig. \ref{T-n} indicates regimes corresponding to
these two phases as well as the two-phase coexistence region and the
continuous transitions line. In this respect it resembles phase
diagrams for some models of multicomponent membranes \cite{TWRL01}.

\bigskip

\noindent \textbf{Acknowledgements.}

\noindent One of us (B.R.) would like to thank Reinhard Lipowsky for 
hospitality and stimulating interactions and Thomas Weikl for helpful
discussions. This work was partially supported by the Committee for
Scientific Research grant 2PO3B 008 23.

\end{document}